\documentclass[12pt]{article}

\begin{document}

\textheight 8.5in
\textwidth 6.5in
\topmargin -30pt

\def\double{\baselineskip 18pt \lineskip 10pt}
\setcounter{page}{0}
\thispagestyle{empty}

\renewcommand{\arraystretch}{1.5}
\newcommand{\be}{\begin{equation}}
\newcommand{\ee}{\end{equation}}
\newcommand{\bea}{\begin{eqnarray}}
\newcommand{\eea}{\end{eqnarray}}
\def\Tr{\mathop{\rm Tr}\nolimits}
\def\Z{{\bf Z}}
\def\R{{\bf R}}
\def\M{{\cal M}}
\def\L{{\cal L}}
\def\N{{\cal N}}

\font\cmss=cmss10 \font\cmsss=cmss10 at 7pt
\def\pp{\partial}
\def\uu{^}
\def\ll{_}
\def\a{\alpha}
\def\b{\beta}
\def\s{\sigma}
\def\g{\gamma}
\def\st{^*}
\def\pr{^\prime}
\def\e{\epsilon}
\def\th{\theta}
\def\thb{\bar{\theta}}
\def\psb{\bar{\psi}}
\def\adot{{\dot{\alpha}}}
\def\bdot{{\dot{\beta}}}
\def\gdot{{\dot{\gamma}}}
\def\dag{^\dagger}
\def\x{\chi}
\def\m{\mu}
\def\n{\nu}
\def\d{\delta}
\def\dbar{\bar{D}}
\def\ebar{\bar{\eta}}
\def\sqd{^2}
\def\hf{{1\over 2}}
\def\hsk{\hskip .5in}
\def\eps{\epsilon}
\def\Ad{\dot{A}}
\def\Add{\hbox{\it \"{A}}}
\def\Phb{\bar{\Phi}}
\def\ibar{\bar{i}}
\def\jbar{\bar{j}}
\def\zb{{\bar{z}}}
\def\baro{\bar{1}}
\def\bart{\bar{2}}
\def\p{\pi}
\def\exp#1{{\rm exp}\left(#1\right)}
\def\oh{\hat{O}}
\def\r{\rho}
\def\t{\tau}
\def\sq{\xi}
\def\xd{\dot{x}}
\def\re#1{{\rm Re}(#1)}
\def\im#1{{\rm Im}(#1)}
\def\psd{\dot{\psi}}
\def\xdd{\hbox{\"{x}}}
\def\k{\kappa}
\def\br#1{\langle#1|}
\def\ip#1#2{\langle#1|#2\rangle}
\def\onep{1_{\rm Pauli}}
\def\trp{{\rm Tr}_{\rm Pauli}}
\def\dgs{^{\dagger 2}}
\def\G{\Gamma}
\def\Xdd{\hbox{\"{X}}}
\def\Xd{\dot{X}}
\def\chd{\dot{\chi}}
\def\ns{ \not \hskip -3pt n  }
\def\phd{\dot{\phi}}
\def\chd{\dot{\chi}}
\def\L{\Lambda}
\def\S{\Sigma}
\def\Sd{\dot{\Sigma}}
\def\Ld{\dot{\Lambda}}
\def\Dpp{D_{++}}
\def\Dmm{D_{--}}
\def\1dag{^{1\dagger}}
\def\2dag{^{2\dagger}}
\def\Qb{{\bar{Q}}}
\def\chb{\bar{\chi}}
\def\O{\Omega}
\def\Rb{\bar{R}}
\def\Sb{\bar{S}}
\def\pb{{\bar{\partial}}}
\def\qb{{\bar{q}}}
\def\rb{{\bar{r}}}
\def\sb{{\bar{s}}}
\def\tb{{\bar{t}}}
\def\ub{{\bar{u}}}
\def\vb{{\bar{v}}}
\def\wb{{\bar{w}}}
\def\ib{{\bar{i}}}
\def\jb{{\bar{j}}}
\def\kb{{\bar{k}}}
\def\lb{{\bar{l}}}
\def\mb{{\bar{m}}}
\def\nb{{\bar{n}}}
\def\P{\Pi}
\def\l{\lambda}
\def\ddot{{\dot{\delta}}}
\def\o{\omega}
\def\R#1#2#3{{{R_{#1}}^{#2}}_{#3}}
\def\vn{\vec{\nabla}}
\def\phb{\bar{\phi}}
\def\Wb{\bar{W}}
\def\lsq{\left [}
\def\rsq{\right ]}
\def\lrd{\left (}
\def\rrd{\right )}
\def\lcb{\left \{}
\def\rcb{\right \}}
\def\ot{\otimes}
\def\o#1#2#3{{{\omega_{#1}}^{#2}}_{#3}}
\def\nt{\tilde{\nabla}}
\def\At{\tilde{A}}
\def\Db{{\bar{D}}}
\def\fb{\bar{f}}
\def\Mb{\bar{M}}
\def\Nb{\bar{N}}

\newcommand{\PSbox}[3]{\mbox{\rule{0in}{#3}\includegraphics{#1}\hspace{#2}}}

\begin{center}
{\large{\bf BPS Domain Wall Junctions in Infinitely
Large Extra Dimensions}}

\vspace*{0.3in}
Sean M. Carroll$^{1}$, Simeon Hellerman$^{2}$, and Mark Trodden$^{3}$
\vspace*{0.3in}

\it
$^{1}$Enrico Fermi Institute and Department of Physics \\
University of Chicago\\
5640 S.~Ellis Avenue, Chicago, IL~60637, USA\\
{\tt carroll@theory.uchicago.edu} \\
\vspace*{0.2in}

$^2$Department of Physics \\
University of California \\
Santa Barbara, California 93106, USA\\
{\tt sheller@physics.ucsb.edu}\\
\vspace*{0.2in}

$^3$Department of Physics \\
Case Western Reserve University \\
10900 Euclid Avenue \\
Cleveland, OH 44106-7079, USA\\
{\tt trodden@huxley.physics.cwru.edu}

\end{center}
\vfill

\begin{abstract}
We consider models of scalar fields coupled to gravity which are
higher-dimensional generalizations of four dimensional
supergravity. We use these models to describe domain wall
junctions in an anti-de~Sitter background.  We derive Bogomolnyi
equations for the scalar fields from which the walls are constructed
and for the metric.  From these equations a BPS-like formula for the
junction energy can be derived.  We demonstrate that such junctions 
localize gravity in the presence of more than one uncompactified 
extra dimension.
\end{abstract}

\vfill

\noindent EFI 99-44

\noindent CWRU-P33-99\hfill hep-th/9911083

\eject

\baselineskip 17pt plus 2pt minus 2pt

\section{Introduction}
\label{introduction}

One of the very few robust predictions of string theory is
the existence of extra dimensions of spacetime.  The apparent
conflict between this prediction and experiment is traditionally
resolved by postulating a direct product structure in 
which the additional spatial dimensions describe a compact
space of Planckian scale.  While this view leads to a 
potentially rich phenomenology, the difficulty in choosing
amongst compactification scenarios and understanding their 
cosmological evolution encourages the search for alternative
pictures.

It has recently become widely recognized that extra dimensions can be
much larger than the Planck scale if the non-gravitational fields of
the Standard Model are confined to a (3+1)-dimensional brane
\cite{add} (see \cite{precursors} for precursors).  Taking this idea
to its extreme, Randall and Sundrum \cite{rs} have shown that gravity
can be effectively localized to a flat four-dimensional submanifold
with a single infinite extra dimension if the five-dimensional bulk
geometry is locally anti-de~Sitter (AdS); see also \cite{visser,gog}.  
It is more difficult, however, to make this idea work for more than one
extra dimension.  Arkani-Hamed {\it et al.}\ \cite{addk} have
suggested that this obstacle can be overcome by considering $n$
individual branes, each of worldvolume dimension $2+n$, in a
$(4+n)$-dimensional spacetime.  If these branes intersect at a
four-dimensional subspace, with the solution in the bulk consisting of
$2^n$ patches of $(4+n)$-dimensional anti-de~Sitter space, 
gravity can be localized to the four-dimensional intersection. 
This scenario has been further elaborated on in 
\cite{csakishirman,nelson},
where the importance of determining the tension associated with the
junction itself was emphasized.

In a brane-world picture, the branes themselves may be either
fundamental objects or solitons (domain walls) constructed 
from fundamental fields.  Supersymmetric domain walls have been
extensively studied (see {\it e.g.} \cite{ds}); in supersymmetric
theories with multiple discrete vacua, there
are typically BPS domain wall solutions which preserve half of
the underlying supercharges.  In (3+1) dimensions, domain wall
junctions have been studied \cite{fmvw,at,tv,gt,cht}, and
it has been shown that there can be a nonzero tension associated
with the junction.  In \cite{gt,cht}, it was argued that 
wall junctions can be BPS states, preserving a single
Hermitian supercharge (of the four in an $N=1$ theory).

We are therefore interested in studying (3+1)-dimensional junctions of
domain walls in higher dimensional supersymmetric theories, coupled to 
gravity.  Some general
properties of supergravity domain walls are well established
\cite{cveticsoleng}.  The simplest context in which (3+1)-dimensional
junctions with $N=1$ SUSY in $4D$
can arise is in six-dimensional theories such as gauged
six-dimensional supergravity with sixteen supercharges, coupled to
matter multiplets.  Unfortunately this set of theories has not yet been
constructed.  However, it has recently been pointed out that a
general class of theories of scalar fields coupled to gravity, with
potentials $V(\phi^i)$ derived from a ``superpotential'', give rise to
classical Bogomolnyi bounds for the tensions of domain walls,
independently of whether or not they can be derived from a specific
supergravity Lagrangian \cite{st,dfgk}.  We therefore
consider models in this class and derive a similar bound for the
tension of domain wall junctions, bypassing the construction of the
supergravity Lagrangians in which they may or may not be embedded.
While BPS domain walls in supergravity are generally singular, in this 
paper we calculate the contribution to the energy of the nonsingular region, 
leaving aside the question of resolving the singularities.

\section{Setup}
\label{setup}

In this section we describe how to adapt the construction of 
Skenderis and Townsend \cite{st} and DeWolfe, Freedman, Gubser and Karch
\cite{dfgk}, describing domain walls in supergravity-like theories,
to a context appropriate for wall junctions.

We consider a set of complex
scalar fields $\{\phi^i\}$ coupled to gravity in $D$ 
spacetime dimensions,
with an action (in units where $\kappa_D^2\equiv 8\pi G_D =1$)
\be
  S = \int d^Dx\, \left({1\over 2}\sqrt{|g|} R 
  + {\cal L}\ll{matter}\right) \ ,
\ee
where the matter Lagrange density is
\be
  {\cal L}\ll{matter} =  \sqrt{|g|}\left[- {1\over 2} K\ll{i\jb} g\uu{ab} 
  (\partial\ll a \phi^i) 
  (\partial\ll b \bar\phi^j) - V(\phi,\bar\phi)\right] \ .
\ee
We have assumed that the number of real scalars is even, allowing
us to group them into complex fields, and that the metric on
field space has the K\"ahler form
$K\ll{i\jb} = \partial^2K/\partial\phi^i\partial\bar\phi^j$,
with $K(\phi,\bar\phi)$ the K\"ahler potential.
We will assume further that the potential $V(\phi,\bar\phi)$ 
can be written in terms of a holomorphic function $W(\phi)$
as
\be
  V = \left(\frac{D-2}{4}\right)\exp{K}\lsq (D-2) 
  K\uu{i\jb}W\ll{;i}\Wb\ll{;\jb} - 2(D-1)W\Wb \rsq,
  \label{potential}
\ee
where
\be
  W\ll{;i} \equiv {{\partial W}\over{\partial\phi^i}} + 
  {{\partial K}\over{\partial\phi^i}}W\ .
\ee
For obvious reasons we will refer to $W$ as the ``superpotential'',
even though our higher-dimensional examples will not be based
on a well-defined supergravity theory.  For the case of trivial
K\"ahler metric, the form (\ref{potential}) was shown to be
necessary for vacuum stability about an AdS background
\cite{townsend84}.

Since our ultimate
objective is to construct domain wall junctions rather than the
individual walls, we consider an
ansatz for the fields which allows for dependence on two spatial
dimensions, $x^{D-2}$ and $x^{D-1}$, which we combine into complex coordinates:
\bea
  z & \equiv & x^{D-2} + ix^{D-1}  \ ,\nonumber \\ 
  \zb & \equiv & x^{D-2} - ix^{D-1}  \ .
\eea
Our metric ansatz in these coordinates is
\be
  ds^2 = e^{2A(z,\zb)}\eta_{\mu\nu}dx^\mu dx^\nu
  + {1\over 2}e^{2B(z,\zb)}\left(dzd\zb + d\zb dz\right)\ ,
  \label{metric}
\ee
where $\m = 0,1,2,\cdots,D-3$.  We assume in addition that the
scalar fields depend only on the extra dimensions,
$\pp\ll\m \phi^i = 0$.  This ansatz is the most general
compatible with $(D-2)$-dimensional Poincare invariance. 

Skenderis and Townsend \cite{st} consider a theory with a single
scalar field.  They study a domain-wall ansatz in which the field only
depends on one spatial coordinate $r$, and derive ``BPS equations'' of
the form $B_I=0$ (with $B_I$ first-order in derivatives of the fields)
by showing that the energy functional can be written in the form 
\be
  E = \int_{-\infty}^{\infty} dr\, \left[(B_1)^2 - (B_2)^2 \right] 
  + {\rm surface~terms}\ .  \label{stenergy} 
\ee 
Such a functional will have a
saddle point (although not necessarily a minimum, due to the minus
sign) at configurations with $B_I=0$, 
which one can verify also solve the full equations
of motion of the theory.  DeWolfe {\it et al.} \cite{dfgk}, meanwhile,
use an equivalent set of equations similar to those
derived from a five-dimensional
supergravity theory in \cite{fgpw}.  In supergravity, one can derive
BPS equations by setting to zero the supersymmetry transformations
generated by certain spinors $\zeta_\alpha$, representing the fact
that BPS states leave some supersymmetry unbroken \cite{cvetic}.  
DeWolfe {\it et
al.} point out that, in the context of scalar fields coupled to
gravity, solutions to such equations continue to solve the full
equations of motion whether or not the superpotential derives from
supergravity.

The additional complication introduced by our extra degrees of freedom
and extra spatial dimension renders the procedure of searching
for BPS equations by writing the
energy functional in a form analogous to (\ref{stenergy})
prohibitively unwieldy. Instead, we would like to employ supergravity
transformations to obtain the BPS equations. However, an appropriate
six-dimensional supergravity theory from which we could derive
BPS-like equations has not been constructed.
What we can do is to first derive
the equations appropriate to ordinary $D=4$ supergravity
by setting supersymmetry variations to zero, 
then generalize those equations to higher dimensions.
Our strategy for deriving BPS equations is thus as follows:
\begin{itemize}
  \item{} Posit the ansatz for the metric and scalar fields 
  appropriate to wall junctions.
  \item{} Derive first-order BPS equations in $D=4$ supergravity
  by setting to zero the supersymmetry transformations of the
  matter fermions $\psi^i_\alpha$ and the gravitino $\psi\ll{a\a}$,
  for some spinor parameter $\zeta_\alpha$.
  \item{} Generalize these equations to higher dimensions
  by introducing constant coefficients, and determining their
  values by the requirement that solutions to
  the BPS equations also solve the equations of motion of a
  theory of scalars coupled to gravity (which may or may not be 
  derivable from supergravity).
\end{itemize}

Once the BPS equations are found, they may be used to
calculate the tension of a junction satisfying them (even in the
absence of specific solutions).  Following \cite{bc}, it is
very plausible that any such solutions are singular in the
cores of the walls.  However, we expect that the appearance
of such a singularity may under favorable conditions
be resolved in a more complete theory (see
for example \cite{baksfet},\cite{fgpw2},\cite{cglp}), and that our
results about the existence of BPS configurations and the
junction tension should still hold.

\section{BPS Equations for Junctions}
\label{bps}

We begin with a derivation of the BPS equations in $D=4$ supergravity
for our ansatz (\ref{metric}).

For calculational convenience we represent the gravitino $\psi\ll{a\a}$
and the matter fermions $\psi\ll\a\uu i$
as Weyl spinors, and the supersymmetry transformation parameter
$\zeta_\alpha$ as a Majorana spinor.  
The supersymmetry transformations for the matter fermions
are then
\be
  \d \psi\ll\a\uu i  =  (P\ll +\G\uu a)\ll{\a\b} \zeta\ll\b \pp\ll 
  a\phi\uu i - \exp{{K \over 2}}
  K\uu{i\jb} \Wb\ll{;\jb} P\ll{+\a\b} \zeta\ll\b \ ,
  \label{mattertx}
\ee
and for the gravitino
\be
  \d\psi\ll{a\a}  =  P\ll{+\a\b} (\nt\ll a\zeta)\ll\b + {1\over 2}
  (P\ll +\G\ll a)\ll{\a\b} \zeta\ll\b \exp{{{K}\over 2}} W \ ,
  \label{gravitinotx}
\ee
where
\be
  \nt\ll a\equiv \nabla\ll a + {{i}\over 2} \G \At\ll a\ ,
\ee
$\nabla\ll a$ is the spacetime covariant derivative acting on spinors,
and we have introduced the projection matrices
$P\ll{\pm} = (1/2)(1\pm\Gamma)$.  Here $\G =-i\G^0\G^1\G^2\G^3$ is 
the four-dimensional chirality matrix, and 
$\At\ll a$ is an auxiliary axial-vector gauge field defined by 
\be
  \At\ll a \equiv {\rm Im} \left( K\ll{,i} \pp\ll a\phi\uu i \right)\ .
  \label{avgf}
\ee

In presenting the supersymmetry transformations 
(\ref{mattertx}) and (\ref{gravitinotx}) we have not chosen a
K\"ahler gauge, thereby preserving 
covariance under a K\"ahler transformation 
in which we assign the matter fermions, the gravitino, and
the local SUSY parameter a nontrivial transformation law:
\bea
  K & \to & K + f(\phi) + \bar{f}(\phb) \nonumber \\
  W & \to & \exp{ - f(\phi)} W \nonumber \\
  \psi\uu i \ll\a & \to & \exp {+{{i}\over 2} {\rm Im} [f(\phi)]}
  \psi\ll\a\uu i  \label{kahlertx} \\
  \psi\ll{a\a} & \to & \exp { - {{i}\over 2} {\rm Im} [f(\phi)]}
  \psi\ll{a\a} \nonumber \\
  \zeta\ll\a & \to & \exp { - {{i}\over 2} {\rm Im} [f(\phi)]\G }
  \ll{\a\b} \zeta\ll\b \ .\nonumber 
\eea
We have chosen the auxiliary axial gauge field 
(\ref{avgf}) in such a way that
under this transformation the covariant derivative of the
SUSY parameter transforms in a linear way:
\bea
  \At\ll a & \to & \At\ll a + \pp\ll a \lcb {\rm Im} 
  [f(\phi)]  \rcb \nonumber \\
  \nt\ll a & \to & \exp{ - {i\over 2} {\rm Im} [f(\phi)] \G }
  \cdot \nt\ll a \cdot \exp{ + {i\over 2} {\rm Im} [f(\phi)] \G }
  \\
  \nt\ll a \zeta & \to & \exp{ - {i\over 2} {\rm Im} [f(\phi)] \G }
  \cdot (\nt\ll a \zeta) \nonumber
\eea
Maintaining this invariance explicitly is useful in deriving
the higher-dimensional analogues of the $D=4$ BPS equations.

We now look for a set of equations corresponding to  a 
$1/4$-BPS configuration, i.e., one which is invariant under a single 
spinor $\zeta$.  These equations will describe the junction states that we
seek. 

The first BPS equation comes from setting the supersymmetric
variation of the matter fermions to zero, $\d \psi\ll\a\uu i =0$. This 
condition is
\be
  (P\ll +\G\uu a)\ll{\a\b} \zeta\ll\b \pp\ll a\phi\uu i 
  - \exp{{{K}\over 2}}
  K\uu{i\jb} \Wb\ll{;\jb} P\ll{+\a\b} \zeta\ll\b = 0\ ,
  \label{deltamatter}
\ee
where $\G\uu a \equiv e\uu a \ll A \G\uu A$ and $\G\ll a \equiv
e\uu A\ll a \G\ll A$, with $A,B,...$ being tangent space indices,
and we have chosen vielbeins $e^a_\m = \exp{-A}\delta^a_\m$,
$e^z_2 = e^\zb_2 = -i e^z_3 = i e^\zb_3 = \exp{-B}$.
For simplicity we will choose the spinor $\zeta$ so
that the particular linear combination of the resulting conditions that we 
obtain is an equation for $\pp\phi$ (and $\pb{\bar \phi}$). This implies
that
\be
  \zeta = \lrd \matrix { g \cr 0 \cr g\uu * \cr 0 }\rrd \ ,
\ee
where $g$ is a function of $z$ and $\zb$, and we are working
in a spinor basis where the $\Gamma$-matrices take the form
\bea
  \Gamma^0 &=& i\sigma^2 \otimes \sigma^1 \cr
  \Gamma^1 &=& \sigma^1 \otimes \sigma^1 \cr
  \Gamma^2 &=& \sigma^3 \otimes \sigma^1 \cr
  \Gamma^3 &=& 1 \otimes \sigma^2 \ .
\eea

It is also convenient to introduce
\be
  \widehat{\G} \equiv \G\uu {2} + i \G\uu{3}
  = \lsq \matrix { 0 & 0 & 2 & 0 \cr 
  0 & 0 & 0 & 0 \cr 0 & 0 & 0 & 0 \cr 0 & -2 & 0 & 0  } \rsq\ ,
\ee
implying
\be
  P\ll + \widehat{\G} = \lsq \matrix { 0 & 0 & 2 & 0 \cr 0 & 0 & 0 & 0 
  \cr 0 & 0 & 0 & 0 \cr 0 & 0 & 0 & 0 } \rsq \ .
\ee
With these conventions, and the assumption
that the solution respects $2$-dimensional Poincare 
invariance, we find that the first term in (\ref{deltamatter}) can be 
written as
\bea
  e\ll A\uu a(P\ll + \G\uu A \zeta)\pp\ll a\phi
  \uu i &\!\!\!\! = &\!\!\!\!
  e\ll A\uu z (P\ll + \G\uu A \zeta) \pp\phi\uu i \nonumber \\
  &\!\!\!\! = &\!\!\!\! 2 \exp{-B} (\pp\phi\uu i ) \cdot 
  \lsq \matrix { 0 & 0 & 1 & 0 \cr 0 & 0 & 0 & 0 
  \cr 0 & 0 & 0 & 0 \cr 0 & 0 & 0 & 0 } \rsq \cdot \lrd \matrix {
  g \cr 0 \cr g\uu * \cr 0 } \rrd \nonumber \\
  &\!\!\!\! = &\!\!\!\! 2 \exp{-B} (\pp\phi\uu i ) \cdot \lrd \matrix {
  g\uu *  \cr 0 \cr 0 \cr 0 } \rrd \ .
  \label{BPS1part1}
\eea
The other part of $\d\psi\uu i \ll\a$ is equal to
\bea
  -  \exp{{K\over 2}} K\uu{i\jb} \Wb\ll{;\jb} 
  P\ll + \zeta
  &\!\!\!\! = &\!\!\!\! -  \exp{{K\over 2}} K\uu{i\jb} \Wb\ll{;\jb} 
  \lsq \matrix { 1 & 0 & 0 & 0 \cr 0 & 1 & 0 & 0 \cr 0 & 0 & 0 & 0 \cr
  0 & 0 & 0 & 0 } \rsq 
  \lrd \matrix {g \cr 0 \cr g\st \cr 0 } \rrd \nonumber \\
  &\!\!\!\! = &\!\!\!\! -  \exp{{K\over 2}} K\uu{i\jb} \Wb\ll{;\jb} 
  \lrd \matrix {g \cr 0 \cr 0  \cr 0 } \rrd
  \label{BPS1part2}
\eea
Therefore, adding (\ref{BPS1part1}) to (\ref{BPS1part2}) and equating
to zero, the relevant BPS equation becomes
\be
  \pp\phi\uu i = {1\over 2} \left({g\over{g\st}}\right)
  \exp{B + {K\over 2}} K\uu{i\jb} \Wb\ll{;\jb} \ .
  \label{B1}
\ee

We now move on to the BPS equations coming from the 
SUSY variation of the gravitino $\psi\ll{a\a}$. 
Requiring $\d\psi\ll{\m\a} =0$ implies
\be
  (P\ll + \nt\ll\m\zeta)\ll\a + {1\over 2}
  (P\ll + \G\ll\m \zeta)\ll\a \exp{{{K}\over 2}} W = 0 \ .
\ee
With $\At\ll\m = \pp\ll\m\zeta = 0$ (since $\mu$ runs from $0$
to $1$), this becomes
\be
  {1\over 4} \omega\ll{\m AB} (P\ll +\G\uu{AB})\ll{\a\b}\zeta\ll\b
   + {1\over 2} 
  (P\ll + \G\ll\m \zeta)\ll\a \exp{{K \over 2}} W \equiv T_1+T_2 = 0\ ,
\ee
where $\G\uu{AB}\equiv [\G^A, \G^B]/2$.
The first term, $T_1$, can be shown to be  
\be
  T_1=  {1\over 2} \eta\ll{\m p} \exp{A - B} \pb A 
  (\G\uu p \widehat{\G} P\ll + \zeta )\ll \a \ 
  \label{BE2part1}
\ee
(with $p=0,1$), for a spinor chosen as above.
The second term, $T_2$, is equal to 
\bea
T_2 &\!\!\!\! = &\!\!\!\!  
{1\over 2} (P\ll + \G\ll\m \zeta)\ll\a \exp{{{K}\over 2}} W 
\nonumber \\
  &\!\!\!\! = &\!\!\!\! 
{1\over 2} \exp{{{K}\over 2}} W e\ll{ p\m} (\G\uu p P\ll -
  \zeta) \ll\a  \nonumber \\
  &\!\!\!\! = &\!\!\!\! {1\over 2} \exp{{{K}\over 2}} W \exp{A} \eta\ll{p\m} 
  (\G\uu p P\ll - \zeta) \ll\a  \ .
  \label{BE2part2}
\eea
Therefore, adding (\ref{BE2part1}) to (\ref{BE2part2}) and setting the sum to 
zero, we obtain that the SUSY variation of $\psi\ll{\m\a}$ vanishes if and 
only if
\be
  \exp{-B} \pb A \widehat{\G} P\ll + \zeta + \exp{{{K}\over 2}} W
  P\ll - \zeta = 0 \ .
\ee
In components, this equation reads
\be
  \exp{-B} ( \pb A ) \cdot 
  \lsq \matrix { 0 & 0 & 0 & 0 \cr 0 & 0 & 0 & -2 \cr
  2 & 0 & 0 & 0 \cr 0 & 0 & 0 & 0 } \rsq \cdot \lrd \matrix
  { g \cr 0 \cr 0 \cr 0 } \rrd + \exp{{{K}\over 2}}
  W  \cdot \lrd \matrix
  { 0 \cr 0 \cr g\st  \cr 0 } \rrd = 0 \ ,
\ee
which yields the second BPS equation:
\be
  \pb A = - {1\over 2}\left({{g\st}\over g}\right)
  \exp{B + {{K}\over 2}} W \ .
  \label{B2}
\ee

In a similar manner, using the combined variations of
$\psi\ll{z\a}$ and $\d\psi\ll{\zb \a}$, we obtain the condition
\be
  g = \exp{\frac{A - i J}{2}}\ ,
\ee
where $J$ is some real function, and a third BPS equation. Writing
\be
  N\equiv B+\frac{K}{2}+iJ\ ,
\ee
the final 4-dimensional BPS equations take the form
\be
  \pp\phi\uu i =  {1\over 2}
  \exp{N\st} K\uu{i\jb} \Wb\ll{;\jb} \ ,
\ee
\be
  \pp A = - {1\over 2}
  \exp{N\st} \Wb \ ,
\ee
\be
  \pp N 
  = {1\over 2} \exp{N\st} (K\uu{i\jb} K\ll{,i} \Wb\ll{;\jb} - \Wb) \ .
\ee

To generalize these BPS equations to higher dimensions, we make the
ansatz that they retain the same form up to potential
dimension-dependent constants, 
while implying the full equations of motion in
a higher-dimensional theory of scalars coupled to gravity.

The independent components of the Ricci tensor for our 
metric are
\bea
  R_{\mu\nu} &\!\!\! = &\!\!\! -4 \eta\ll{\m\n} \exp{2(A-B)} 
  \lsq \pp\pb A + 
  (D-2) (\pp A)(\pb A) \rsq \ , \nonumber \\
  R\ll{zz} &\!\!\! = &\!\!\! (D-2) \lsq 2(\pp A)(\pp B)  
  - (\pp A)\sqd - \pp\sqd A \rsq \ , \\
  R\ll{z\zb} &\!\!\! = &\!\!\! - (D-2) \lsq\pp\pb A + 
  (\pp A)(\pb A) \rsq - 2 \pp\pb B\ .\nonumber
\eea
(The $\zb\zb$ component of $R_{ab}$, like that of the
energy-momentum tensor below, is obtained from the $zz$ component
by replacing $\pp \leftrightarrow \pb$.)
The energy-momentum tensor is
\bea
  T_{\m\n} &\!\!\! = &\!\!\!  
  - 2 \eta\ll{\m\n}\exp{2(A-B)} K\ll{,i\jb} 
  \lsq (\pp\phi\uu i)(\pb\phb\uu j)
  +(\pb\phi\uu i )(\pp\phb\uu j)\rsq - \nonumber \\
  & & \qquad \eta\ll{\m\n}\exp{2A}V(\phi,\phb) \ ,
  \nonumber \\
  T\ll{zz} &\!\!\! = &\!\!\!  
  2 K\ll{i\jb}(\pp\phi\uu i)(\pp\phb\uu j) \ ,  \\
  T\ll{z\zb} &\!\!\! = &\!\!\! - {1\over 2} \exp{2B} V(\phi,\phb) 
  \nonumber \ .
\eea
There are thus three independent Einstein equations; 
the $\mu\nu$ equation gives
\bea
  2(D-3) \pp\pb A + 2\pp\pb B +  {{(D-2)(D-3)}}
  (\pp A)(\pb A)  = \nonumber \\
  K\ll{,i\jb} \lsq (\pp\phi\uu i)(\pb\phb\uu j)
  +(\pb\phi\uu i )(\pp\phb\uu j)\rsq - {1\over 2}\exp{2B}V(\phi,\phb) 
  \label{I} \ ,
\eea
the $zz$ equation gives
\be
  (D-2) \lsq 2(\pp A)(\pp B)  - (\pp A)\sqd - \pp\sqd A \rsq
   = 2 K\ll{i\jb}(\pp\phi\uu i) (\pp\phb\uu j)
  \label{II} \ ,
\ee
and the $z\zb$ equation gives
\be
  (D-2) \lsq \pp\pb A + (D-2)(\pp A)(\pb A) \rsq = 
  - {1\over 2} \exp{2B} V(\phi,\phb)\ .
  \label{III}
\ee
In addition we have the scalar field equation of motion,
\bea
  (D-2) \lsq 2 (\pp A)(\pb\phi\uu i) + 2 (\pb A)(\pp\phi\uu i)\rsq
  + 4 \pp\pb\phi\uu i + \nonumber \\
  +4 K\uu{i\mb}K\ll{,jk\mb}(\pp\phi\uu j)(\pb\phi\uu k)
   = \exp{2B} K\uu{i\jb}V\ll{,\jb}\ .
\label{IV}
\eea
Substitution of the four-dimensional BPS equations, augmented
by appropriate constant coefficients,
back into these $D$-dimensional equations of motion yields the following
$D$-dimensional BPS equations:
\be
  \pp\phi\uu i = {{(D-2)}\over 4}  
  \exp{\kappa_D N\st} K\uu{i\jb} \Wb\ll{;\jb} \ ,
  \label{bps1}
\ee
\be
  \pp A = - {\kappa_D\over 2} 
  \exp{\kappa_D N\st} \Wb \ ,
  \label{bps2}
\ee
\be
  \pp N 
  = {\kappa_D\over 2} \exp{\kappa_D N\st} \left[{(D-2)\over 2} 
    K\uu{i\jb} K\ll{,i} \Wb\ll{;\jb} -  \Wb\right] \ ,
  \label{bps3}
\ee
where we have restored explicit factors of $\kappa_D$ for later
convenience.
Together, these equations imply the equations of motion for gravity in 
$D$ dimensions coupled to a set of complex scalars with K\"ahler 
potential $K$ and potential energy given by (\ref{potential}).

\section{BPS Bounds and the Junction Energy}

In this section we compute the energy of configurations described by the BPS
equations that we have derived.  We are interested in static configurations,
for which the energy density is defined as the negative of the action:
\be
  E=\int dx^{D-1}dx^{D-2}\, (I_{\rm grav}+I_{\rm matter}) \ ,
  \label{energydef}
\ee
where 
\bea
  I_{\rm grav} &\!\!\! \equiv - &\!\!\! {1\over 2} \sqrt {|g|} R 
  \nonumber \\
  &\!\!\! = 2 &\!\!\! \exp{(D-2) A}
  \lsq 2 \pp\pb(B+(D-2)A) \right. \nonumber \\
  & & + \left. (D-1)(D-2)(\pp A)(\pb A) \rsq \ ,
\eea
and
\bea
  I_{\rm matter} &\!\!\! \equiv &\!\!\! 
  \exp{(D-2)A}\left\{2 K\ll{i\jb} (\pp\phi\uu i)(\pb\phb\uu j)
  + 2 K\ll{i\jb} (\pp\phb\uu j)(\pb\phi\uu i)\right. \nonumber \\
  & & + \frac{(D-2)}{4}\exp{2B+K} \lsq
  (D-2) K\uu{i\jb} W\ll{;i} \Wb\ll{;\jb}\right. \nonumber \\
  & & \qquad\qquad \left. \left.
  - 2 (D-1) W \Wb \rsq 
  \right\} \ .
\eea
The gravitational piece of this can be rewritten in the form
\bea
  I_{\rm grav} = &\!\!\! (D-2) &\!\!\! \exp{(D-2)A}
  \left\{(\pp A)(\pb K) +  (\pb A)(\pp K)
   - 2  (\pb A)(\pp N) \right. \nonumber \\
   &\!\!\! -  &\!\!\! \!\!\!\! 2  \left. 
  (\pp A)(\pb N\st) - 2 (D-3) (\pp A)(\pb A)
  \right\} \nonumber \\
  &\!\!\! + &\!\!\! \!\!\!\! \S\ll 1 \ ,
  \label{grav}
\eea
where $\S\ll 1$ is a total derivative given by
\bea
  \S\ll 1 &\!\!\! \equiv &\!\!\! \pp \lsq \exp{(D-2)A}(2 (D-2) \pb A
  + 2 \pb N - \pb K ) \rsq \nonumber \\
  & & +  \pb \lsq  \exp{(D-2)A} (2 (D-2) \pp A
  + 2 \pp N\st
 - \pp K ) \rsq \ .
\eea
Similarly, the matter piece of the energy density can be rewritten as
\bea
  I_{\rm matter} &\!\!\! = &\!\!\! \exp{(D-2)A} \left\{
  4 K\ll{i\jb} (\pp\phi\uu i)(\pb\phb\uu j) \right. \nonumber \\
  & & +  (D-2) (\pp A) \lsq 2  K\ll{,\jb}(\pb\phb\uu j)
  - (\pb K) \rsq 
  + \left. 
  (D-2)(\pb A) \lsq 2 K\ll{,i} (\pp\phi\uu i) - (\pp K)
  \rsq \right. \nonumber \\
  & & +  \left. \frac{(D-2)}{4} \exp{2B+K} \lsq
  (D-2) K\uu{i\jb} W\ll{;i} \Wb\ll{;\jb} - 2
  (D-1) W \Wb \rsq \right\} \nonumber \\
  & & + \S\ll 2 \ ,
  \label{matter}
\eea
where $\S\ll 2$ is a total derivative term given by
\bea
  \S\ll 2 &\!\!\! \equiv &\!\!\! \pp 
  \lsq \exp{(D-2)A}( K\ll{,i} \pb\phi\uu i
  - K\ll{,\jb} \pb\phb\uu j ) \rsq \nonumber \\
  & & +  \pb \lsq \exp{(D-2)A} ( K\ll{,\jb} \pp\phb\uu j
  - K\ll{,i} \pp\phi\uu i ) \rsq \ .
\eea
Adding (\ref{grav}) to (\ref{matter}), the resulting integrand is
\bea
   I\ll{grav} + I\ll{matter} &\!\!\! = &\!\!\! \exp{(D-2)A}\left[
4K\ll{i\jb}(B1)\uu i (B1)\uu{*\jb} 
+ 2(D-2)K\ll{,i} (B1)\uu i (B2)\st \right. \nonumber \\
&\!\!\! + &\!\!\! \left. 2(D-2) K\ll{,\jb} (B2) (B1)\uu{*\jb}
- 2(D-2)(B3)(B2)\st \right. \nonumber \\
&\!\!\! - &\!\!\! \left.2(D-2)(B2) (B3)\st
- 2(D-2)(D-3) (B2) (B2)\st \right] \nonumber \\
&\!\!\! + &\!\!\! \S\ll 1 + \S\ll 2 +\S\ll 3 \ ,
\eea
where each of the terms $(B1)^{i}$, $(B2)$, $(B3)$ vanishes independently by 
virtue of BPS equations (\ref{bps1}), (\ref{bps2}), (\ref{bps3}) respectively, 
and the third total derivative is
\bea
  \S\ll 3 &\!\!\! \equiv &\!\!\!
  (D-2) \pp\lsq \exp{N + (D-2)A} W \rsq \nonumber \\
  & & + (D-2) \pb \lsq \exp{N\st+ (D-2)A} \Wb \rsq \ .
\eea

We therefore find that the total energy as defined by (\ref{energydef}),
can be expressed for a BPS junction state as
\be
  E_{\rm total} = \int d\uu D x \, (\S\ll 1 + \S\ll 2 + \S\ll 3) \ .
  \label{energy}
\ee
To interpret this expression, it is useful to consider the
flat-space limit of BPS junctions, considered (in four dimensions)
in \cite{gt,cht}.  Setting $\kappa_D=0$ in (\ref{bps1})
yields
\be
  \pp\phi\uu i = {{(D-2)}\over 4}  
  K\uu{i\jb} \Wb\ll{,\jb} \ ,
\ee
which is precisely the flat-space BPS equation for wall
junctions.  Meanwhile, in this limit (\ref{bps2}) simply
becomes $\pp A = 0$, so $A$ is a constant that can be set to
zero by a trivial coordinate transformation. Further, 
(\ref{I}) implies that in the $\kappa_D=0$ limit $B$ is a harmonic
real function of $z$ and $\zb$, and therefore can be
brought to zero by a holomorphic coordinate transformation of $z$ and
$\zb$. Thus, we recover flat spacetime in this limit.
However, we also find important information from the 
${\cal O}(\kappa_D^1)$ term in the small-$\kappa_D$ limit;
since the ${\cal O}(\kappa_D^0)$ term in $A$ can be set to
zero, we can define $A' \equiv -A/\kappa_D$, which in this
limit satisfies
\be
  \pp A' = {1\over 2} \Wb \ ,
\ee
which is precisely the equation satisfied by
the preprofile for the superpotential for BPS junctions discussed 
in \cite{cht}.  Thus, in this limit,
the $\S\ll 1$ term in (\ref{energy}) vanishes, while
$\S\ll 3$ becomes the 
sum of the energy of the individual walls and $\S\ll 2$ becomes
the tension associated with the junction itself.
As shown in \cite{oins}, the junction tension is negative.

\section{A graviton zero mode}
\label{graviton}

Arkani-Hamed {\it et al.} \cite{addk} 
have shown that the effective $(D-2)$-dimensional
theory on a junction  of infinitely thin walls
should contain a localized graviton. 
In this section we check that our solitonic walls (of 
finite thickness) also contain a distinguished zero mode
metric fluctuation which, for a well-behaved junction solution, will
be localized to the hub.  We find that there exists a closed form
expression for this mode, and note that its existence
appears to be a general feature of domain wall junctions in AdS, rather
than a special feature that occurs as a result of supersymmetry.  We will
also examine the graviton zero mode for a single domain wall, and make
an explicit connection with the wave equation of \cite{rs} and its
zero-eigenvalue solution in the long-distance limit.

We begin with a set of background fields $\phi\uu i(z,\zb)$ and
background metric $g\ll{ab}$ satisfying the Bogolmonyi equations
(\ref{bps1})-(\ref{bps3}), and consider linearized equations for
the metric perturbations by substituting
$g_{ab}\rightarrow g\ll{ab} + \d g\ll{ab}$ into the equations of
motion (\ref{I})-(\ref{IV}).  We find that there is a
self-consistent solution without introducing perturbations in
the scalar fields.  We take the metric fluctuation
$\d g\ll{ab}$ to be of the form
\be
  \d g\ll{zz} = \d g\ll{\zb\zb} = \d g\ll{z\zb} = \d g\ll{\m z} = 0 \ ,
\ee
\be
\d g\ll{\m\n} = e^{ip\cdot x}e^{2A} \Psi(z,\zb) P\ll{\m\n} \ ,
\ee
where $\Psi(z,\zb)$ is the graviton wavefunction.
Here $p\cdot x \equiv p\ll{\m} x\uu\m $, with $p\ll\m$ some $(D-2)$ vector,
and $P\ll{\m\n}= P\ll{\n\m}$ is a polarization tensor
satisfying
\be
\pp\ll \a(p\ll\n) = \pp\ll\a (P\ll{\n\s} ) =  
 \eta\uu{\m\n} p\ll\m p\ll\n = \eta\uu{\m\n} P\ll{\m\n} = \eta\uu{\m\n}
p\ll{\m} P\ll{\n\s} = 0 \ .
\ee

It is convenient to write the perturbed Einstein equation as
\be
  \delta R_{ab} - \delta \left[{1\over 2} Rg_{ab}
  + T_{ab}\right] = 0\ .
  \label{epert}
\ee
Straightforward computation then shows that the first term,
the variation of the Ricci tensor, is
\begin{eqnarray}
  \d R\ll{\m \n} = & \!\!\!
  - &\!\!\!  e^{ip\cdot x}e^{2A - 2B}
  \left\{ (D-2) (\pp A) (\pb \Psi) + (D-2)(\pb A)(\pp\Psi)\right.
  \nonumber \\
  &\!\!\!\! & \left. + 2 \pp\pb \Psi
  + 4 \lsq \pp\pb A + (D-2) (\pp A)(\pb A) \rsq \Psi \right\}P\ll{\m\n} \ ,
\end{eqnarray}
with all other components vanishing. 
The second term in (\ref{epert}) is
\be
  {1\over 2} \d (R g\ll{\m\n}) + \d T\ll{\m\n} =  
  e^{ip\cdot x} e^{2A} \left({{2 V}\over{D-2}}\right) \Psi P\ll{\m\n}\ .
  \nonumber
\ee
Therefore, the linearized equations of motion are
equivalent to the Schr\"odinger equation
\bea
  \lsq - 2\pp\pb \right. &\!\!\! - &\!\!\!\! \left. 
  (D-2) (\pp A)\pb - (D-2)(\pb A)\pp \rsq \Psi
  \nonumber \\
  &\!\!\! - &\!\!\!\! 
  \lsq 4 \pp\pb A + 4(D-2)(\pp A)(\pb A) 
  + {{2V}\over {D-2}} \rsq \Psi = 0\ .
\eea
However, using equations (\ref{bps1})-(\ref{bps3}) and the definition
(\ref{potential}) of the scalar potential, we see that
the second term vanishes, and
the distinguished solution to the massless Schr\"odinger equation is simply
$\Psi = {\rm constant}$, or
\be
  \delta g_{\m\n} \propto e^{ip\cdot x} e^{2A} P_{\m\n}\ .
  \label{deltag}
\ee

Lacking an explicit solution to the BPS equations, we can
nevertheless check that our graviton fluctuation reduces to
that of \cite{addk} in the long-distance (thin-wall) limit.
This serves to verify that the integral of the square norm of 
the wavefunction is indeed convergent far from the walls.
For the case of a four-wall junction in two extra dimensions,
the metric of \cite{addk} can be written
\be
   ds^2 = {1\over {(k|y_1| + k|y_2| +1)^2}}
  (\eta_{\mu\nu}dx^\mu dx^\nu + dy_1^2 +dy_2^2)\ ,
\ee
where we are using $y_i$ instead of $\bar{z}_i$ to
avoid confusion with our complex coordinates, and $k\equiv
(\sqrt{2}L)^{-1}$, with $L$ the AdS curvature radius in the
bulk.  This is a special case of our metric (\ref{metric}), with
\be
  e^{A} = e^{B} = (k|y_1| + k|y_2| +1)^{-1}\ .
\ee
The associated graviton zero-mode is
\be
  \delta g_{\mu\nu} \propto e^{ipx} (k|y_1| + k|y_2| +1)^{-2}
  P\ll{\m\n}\ ,
\ee
which therefore agrees with the form given by (\ref{deltag}).
Thus, integrating the norm square of the wavefunction $\Psi$ over the
$z$-$\zb$ plane, we find that the norm is finite in this
thin-wall limit.  More precisely, since the
metric may be singular near the walls, we have shown that there is a
distinguished zero-frequency wavefunction whose lack of
normalizability can only arise as a result of the singular behavior of
the solution when curvatures become large and the supergravity
approximation breaks down.  However, since we know that under favorable 
conditions the wall solutions themselves can become nonsingular when
embedded in higher-dimensional supergravity, we believe that the
normalizable zero mode is likely to be a robust feature of the full
supergravity (or string theory) state.

\section{Discussion}
\label{discussion}

The idea that the three observed dimensions of our universe may not be
the only large spatial dimensions, and that our universe may exist as
a distinguished 3-manifold embedded in a higher dimensional space, is
an intriguing alternative to conventional compactification \cite{rs}.
If the true universe is $(4+1)$ dimensional, then the relevant
$3$-manifold is a domain wall, or $3$-brane.  However, for larger
background spaces, the $3$-manifold representing our universe may lie
at an intersection of a number of such codimension one branes
\cite{addk}.  (For alternative proposals supporting more than one
extra dimension, see \cite{cohenkaplan}, \cite{gregory}.)  
In this paper we have
considered models of scalar fields coupled to gravity, which are
higher dimensional generalizations of four dimensional supergravity.
We were able to derive first-order BPS equations appropriate to
junction configurations with $(3+1)$-dimensional Poincare invariance
on the hub.

The question of the cosmological constant as measured by 
inhabitants of the brane world is a crucial one in 
Randall-Sundrum scenarios and their generalizations.
In models where the branes are put in by hand as delta-function
sources, the induced cosmological constant can be tuned to 
zero by appropriately balancing the brane tension against
the bulk cosmological constant (either in the original
single-brane scenario \cite{rs} or in junction models \cite{addk}).
In models where the branes are solitons constructed in a field
theory, there may be additional constraints on the induced
brane geometry.  In the single-brane case, Behrndt and 
Cvetic have argued that a BPS state in an actual supergravity
theory will automatically have a flat induced geometry \cite{bc},
although it may also necessarily be singular.  The examples of
Skenderis and Townsend \cite{st} and DeWolfe {\it et al.} \cite{dfgk}
demonstrate that it is possible to find non-singular BPS-like
solutions with flat induced geometries, 
although not necessarily in theories that are
truncations of supergravity.  

In the case of junctions, there is an additional complication
due to the tension of the junction itself; one might worry that
there could not simultaneously be flat walls, a flat junction,
and a nonvanishing junction tension.  We proceeded
with an ansatz satisfying $(3+1)$-dimensional Poincare invariance,
implying a flat geometry on the junction itself, and were able
to derive a consistent set of BPS-like equations in a theory
of scalars coupled to gravity.
Since our solutions feature a nonzero
junction tension, the external geometry in
the thin-wall limit will therefore resemble the AdS patches of \cite{addk}
along with extra global identifications representing a deficit
angle due to the gravitational influence of the junction, as
discussed in \cite{nelson}.  Our ability to derive the
appropriate BPS equations is good evidence that such solutions
exist (although they may be singular); it would be interesting
to further explore the relationship between the tensions on
the walls and the junction, the bulk cosmological constant, and
the induced geometry.  It is unclear, for example, whether
preserving supersymmetry necessarily induces a flat metric
on a junction.

In an infinitely thin brane junction universe, it has been argued that the
graviton is confined to the junction \cite{addk}. 
Here we have examined this question for junctions of finite width, 
constructed as solitonic solutions to field theories. In this microphysical
context we have shown that the junction state admits a graviton zero mode,
normalizable away from the walls, and thus localizes gravity effectively. 
In addition, this result does not 
appear to depend crucially on supersymmetry. Rather it appears to be
a general feature of domain wall junctions in an
anti-de~Sitter background.

The models considered in this paper are particle-physics realizations of a
generalized Randall-Sundrum scenario, and as such should provide a testing 
ground for details of the theory. We have demonstrated how various aspects
of the brane-world idea arise in the context of these models, and hope that
further investigation will cast light on the implications of this 
scenario for the observable universe.

\section*{Acknowledgments}

We would like to thank Oliver DeWolfe, Dan Freedman,
Jeff Harvey, Nemanja Kaloper, Per Kraus, Rob Meyers, Joe Polchinski,
Walter Polkosnik, and Raman Sundrum for valuable discussions.  
This work was supported in part by the National 
Science Foundation under grants PHY/94-07194 and PHY/97-22022, and by 
the U.S. Department of Energy (D.O.E.)

\end{document}